%% file: arXiv.tex
\documentclass[twoside,a4paper,11pt]{sea22}
\usepackage{graphicx}
\usepackage{hyperref}
\usepackage{movie15}
\usepackage{color}
\usepackage{booktabs} 
\usepackage{siunitx}  
\usepackage{xcolor}
\usepackage{natbib}    
\usepackage[export]{adjustbox}
\usepackage{animate}
\usepackage{pdflscape}
\input{mlaeff}

\input{isolatin.sty}

\newcommand{\HII}{\mbox{H\,{\sc ii}}}
\newcommand{\Teff}{\mbox{$T_{\rm eff}$}}
\newcommand{\logg}{\mbox{$\log g$}}
\newcommand{\vsini}{\mbox{$v\sin i$}}

\newcommand{\VO}[1]{Villafranca~O-{#1}}

\newcommand{\lili}{LiLiMaRlin}
\newcommand{\MTW}{M$^3$W}

\begin{document}
\pagenumbering{arabic}
\pagestyle{myheadings}
\thispagestyle{empty}
\vspace*{-4.0cm}
\textit{\flushleft\small
Highlights of Spanish Astrophysics XIII\\
Proceedings of the XVII Scientific Meeting of the Spanish Astronomical Society \\
\hspace{5mm}held on 13-17 July 2026, in Tarragona, Spain.}
\vspace*{1.0cm}


\begin{flushleft}
{\bf {\LARGE
%
Results on young stellar clusters with the WEAVE LIFU
%
}\\
\vspace*{1cm}
%
J. Maíz Apellániz$^1$, G. Holgado$^{2,3}$, C. Fariña$^{2,4}$, \& S. R. Berlanas$^{1,2,3}$
%
}\\
\vspace*{0.5cm}
%
$^1$ Centro de Astrobiología, CSIC-INTA, Spain\\
$^2$ Instituto de Astrofísica de Canarias, Spain\\
$^3$ Universidad de La Laguna, Spain\\
$^4$ Isaac Newton Group of Telescopes, Spain\\
%
\end{flushleft}
%
\markboth{
Young stellar clusters with the WEAVE LIFU
}{ 
%
Ma{\'\i}z Apell\'aniz et al.
%
}
\thispagestyle{empty}
\vspace*{0.4cm}
\begin{minipage}[l]{0.09\textwidth}
\ 
\end{minipage}
\begin{minipage}[r]{0.9\textwidth}
\vspace{1cm}
\section*{Abstract}{\small
%
We present the first results of two Spanish TAC programs using the WEAVE LIFU that have been obtaining data since 
October 2023. The first one is obtaining high-resolution multi-epoch spectroscopy of the nine stellar clusters that contain a 
minimum number of 3 O-type stars (and many more B stars) within the field of view of the WEAVE LIFU and that are accessible from 
La Palma with the purpose of studying the spectroscopic multiplicity of the OB stars. The second program is obtaining single-shot 
low-resolution spectroscopy of as many stellar clusters with OB stars as possible with the purpose of characterizing the OB-star 
population in those regions that are too crowded to be properly studied with the guaranteed time SCIP program. That program has 
already observed 54 clusters in a wide range of environments, from \HII\ regions to gas-free clusters, and distances,
from clusters in the solar neighbourhood to (nearby) extragalactic systems. We also present the tools developed for this program, 
which include a data cube processing and visualisation utility and UNWIND, a code that fits and
subtracts the ISM signature in the observed spectra.
%
\normalsize}
\end{minipage}
%
%


\section{The WEAVE LIFU}              

$\,\!$\indent The \href{https://www.ing.iac.es/astronomy/instruments/weave/lifu.html}{WEAVE LIFU} provides spectroscopy over an
hexagonal FOV of $90\arcsec\times78\arcsec$ with 547 fibres, each one of them with a diameter of 2\farcs6, plus eight additional 
mini-arrays of seven fibres each for the sky. Each fibre yields one blue and one red spectrum in three possible combinations: 
LR (\num{3660}-\num{6060}~\AA\ + \num{5790}-\num{9590}~\AA\ at $R=\num{2500}$), BHR (\num{4040}-\num{4650}~\AA\ + 
\num{5950}-\num{6850}~\AA\ at $R=\num{10000}$), and GHR (\num{4630}-\num{5450}~\AA\ + \num{5950}-\num{6850}~\AA\ at 
$R=\num{10000}$), with an inter-CCD wavelength gap for each blue/red range.

The filling factor of the array is 0.55, so a dithering pattern is usually applied, the most common one being a 
3-point pattern. WEAVE OBs are 1-hour long which, including overheads, leads to the most commonly used observation pattern for a
field consisting of a $3\times1020$~s exposure sequence. However, many of our targets include stars brighter than the saturation
limit of 12-13 mag (the exact value dependes on the SED and seeing), with some as bright as 4th magnitude. This led us in
many cases to require shorter exposures times, which have to be accommodated into the 1-hour OB requirement. Therefore, we used
other dithering patterns (some developed by us) with as many as 16 points, which allowed us to improve both the spatial resolution
and the dynamic range of our observations.


\section{The multiple-O cluster program}      

$\,\!$\indent This poster presents results from two observing programs of the Spanish TAC. The first one, started in October 2023,
is observing the nine clusters accessible from La Palma that contain three or more O stars (selected from GOSC, 
\citealt{Maizetal04b}) within the LIFU FOV (one of them, \VO{005}, has not been observed yet as we are waiting for the instrument team
to observe farther south than now). All clusters contain additional B stars. The goal of the program is to
study the massive-star multiplicity in those clusters as part of \MTW\ \citep{Maizetal26b} by combining the spectra with additional
ones from \lili\ \citep{Maizetal19a} and, eventually, with further \textit{Gaia} data. Targets are observed once in both LR and BHR
and multiple times in GHR to (a) identify which stars are spectroscopically single and (b) derive the spectroscopic orbits of the
ones that are multiple. The multiple-O cluster sample is given in Table~\ref{clustersample}.

\begin{table}
\label{clustersample}
\caption{Multiple-O cluster program sample.}
\centerline{
\begin{tabular}{lcccl}
\midrule
Cluster         & RA         & dec           & \HII? & Sample O star     \\
\midrule
\VO{005}        & 17:24:42.5 & $-$34:12:14.4 & yes   & Pismis~24-1~A     \\
HD~\num{167834} & 18:17:32.9 & $-$12:06:10.8 & no    & HD~\num{167834}~A \\
NGC~\num{6604}  & 18:18:04.6 & $-$12:14:24.0 & no    & MY~Ser~Aa,Ab      \\
\VO{009}        & 18:20:32.2 & $-$16:10:30.0 & yes   & GLS~\num{19618}   \\
\VO{007}        & 20:33:10.3 & $+$41:13:12.0 & no    & Cyg~OB2-22~A      \\
\VO{008}        & 20:33:16.3 & $+$41:18:57.6 & no    & Cyg~OB2-8~A       \\
BD~+55~2722     & 22:19:01.4 & $+$56:07:26.4 & weak  & BD~+55~2722~A     \\
Pacman~nebula   & 00:52:47.5 & $+$56:37:26.4 & yes   & HD~\num{5005}~A   \\
BD~+00~1617     & 06:48:49.4 & $+$00:22:37.2 & no    & BD~+00~1617~B     \\
\midrule
\end{tabular}
}
\end{table}


\section{The cluster and H II region survey program}      

$\,\!$\indent The second Spanish TAC program was designed as a complement to the SCIP survey of massive stars in the
northern Galactic Plane with the WEAVE MOS. Due to its fibre positioning limitations, the MOS 
cannot observe all of the stars in crowded cluster cores. Furthermore, some of the brightest stars in clusters saturate the
standard $3\times1020$~s exposures and in other cases clusters are immersed in \HII\ regions, requiring nearby fibres to subtract
the nebular emission. A LIFU program solves those issues.

We used the ALS catalog \citep{Pantetal25b} to identify the sample, which was expanded with clusters and \HII\ 
regions located outside the Galactic Plane and with extragalactic \HII\ regions that could be reached with WHT. Currently we have 
observed 54 targets in LR, with the number possibly reaching 100 by 2027 if the Spanish TAC grants the project additional time.

\begin{figure}
\centerline{
 \includegraphics*[width=\linewidth]{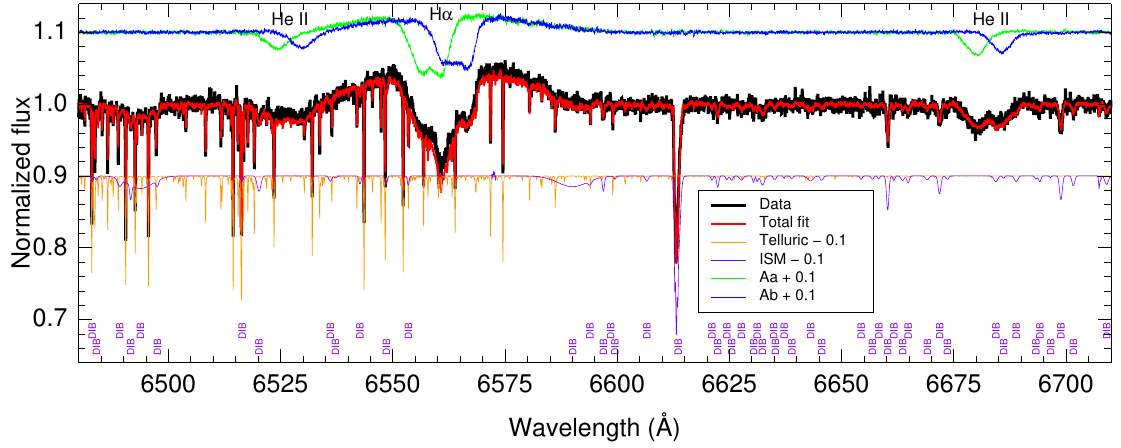}
 }
\caption{Example of UNWIND output for the H$\alpha$ region of a GLS~\num{11448}~Aa,Ab high-resolution spectroscopic epoch and the 
         result of combining 77 epochs. The top spectra (green and blue) show the normalized output for the two components, shifted 
         horizontally to the RVs of the epoch and upwards 0.1 continuum units, and diluted by their flux fractions. The bottom 
         (orange and purple) spectra show the fitted telluric lines, specific to this epoch, and ISM spectrum, mostly DIBs and common
         to all epochs, in the two cases shifted downwards 0.1 units. The central spectra show the data (black) and fit (red) for 
         this epoch, the fit being the sum of the green and blue spectra (minus one to leave the continuum fixed) multiplied by the 
         orange and purple spectra.}
\label{fig1}   
\end{figure}


\section{Science objectives}      

$\,\!$\indent The main immediate goal of the two programs is to extract as many spectra of cluster members as possible, eliminating 
both the nebular contamination by PSF fitting to the data cube and the contributions from telluric lines and the ISM in absorption 
using the UNWIND code (see below), and recalibrating in flux using \textit{Gaia} spectrophotometry. As part of the process, the 3-D 
cube of nebular emission will also be obtained.

\textbf{Stellar analysis.} The stellar spectra will be classified using MGB \citep{Maizetal12} and the latest 
classification grid from the GOSSS project \citep{Maizetal11}. The stellar parameters (\Teff, \logg, \vsini\ldots) will be fitted 
by comparing them with FASTWIND models \citep{Pulsetal05} using the techniques of \citet{Holgetal18}.

\textbf{Stellar multiplicity.} The high-resolution data from the multiple-O cluster program will be integrated with the
rest of the data from the \MTW\ program. For the O stars we have independent high-resolution multi-epoch spectroscopy and the
combination will be used to obtain spectroscopic orbits in an analogous way to our previous work in e.g. \citet{Holgetal25a} and
\citet{Barbetal26}. For the B stars we will follow a similar strategy using the WEAVE spectra alone. The ultimate goal of \MTW\ is to 
obtain the most complete catalog (hundreds of systems) of massive-star orbits and, from there, perform a complete study of massive
stellar multiplicity in the solar neighbourhood.

\textbf{Cluster analysis.} The clusters in both surveys will be added to the Villafranca catalog of Galactic OB-star 
groups \citep{Maizetal20b,Maizetal22a,Maizetal25}, noting that some are already included there. The newly obtained spectroscopy will 
be combined with \textit{Gaia}~DR4 data to derive improved distances and cluster memberships. In combination with the multiplicity 
analysis (see previous paragraph), the information will be used to derive the IMF for the clusters.

\textbf{Extinction and the ISM.} Even though the primary purpose of these projects is to analyse the stars, the ISM will
also be studied. Extinction will be studied by [a] fitting the obtained SEDs and 2MASS photometry to obtain the amount and 
type of extinction \citep{Maizetal14a,MaizBarb18} and [b] comparing the obtained values with those from the nebular H lines when an 
\HII\ region is present \citep{Maizetal98,Maizetal04a}. The ISM extracted with UNWIND will be analysed within the CollDIBs project
\citep{Maiz15a}, with special attention to the ultrawide 7700~\AA\ DIB \citep{Maizetal21a}.


\section{Tools}      

$\,\!$\indent Two new tools have been developed to process the data. The first one is a post-processing pipeline that starts from the
single-fibre spectra delivered by the WEAVE L1 pipeline and forms a 3-D cube in RA+Dec+wavelength. The tool allows for arbitrary
dithering patterns and for adjusting the drizzling parameter \citep{FrucHook02} to account for different samplings and seeings. A PSF
fitting code extracts the individual spectra, which are then recalibrated in flux using \textit{Gaia}~XP spectrophotometry, as some of 
the pipeline versions have issues with e.g. dichroic corrections. Figure~\ref{fig2} shows sample fields processed with the tool at two
different wavelengths along with a \textit{Gaia}~DR3 chart of the field.

The second tool is called UNWIND and is presented in more detail in \citet{Maizetal26b}. The primary use of the tool is 
to deconvolve the components of SB2 systems but here it is also used for the additional purposes of rectification, telluric-line 
elimination, and subtraction of the ISM (ionic, molecular, and DIB) lines prior to the analysis of the stellar spectra. An example of 
the application of UNWIND is shown in Fig.~\ref{fig1}.


\small{\bibliographystyle{aa} 
\bibliography{general}\vspace{0.75in}} 

\begin{landscape}
\begin{figure}
 \centerline{
  \hspace{-3.2cm}
  \raisebox{-8mm}{\includegraphics*[width=0.410\linewidth]{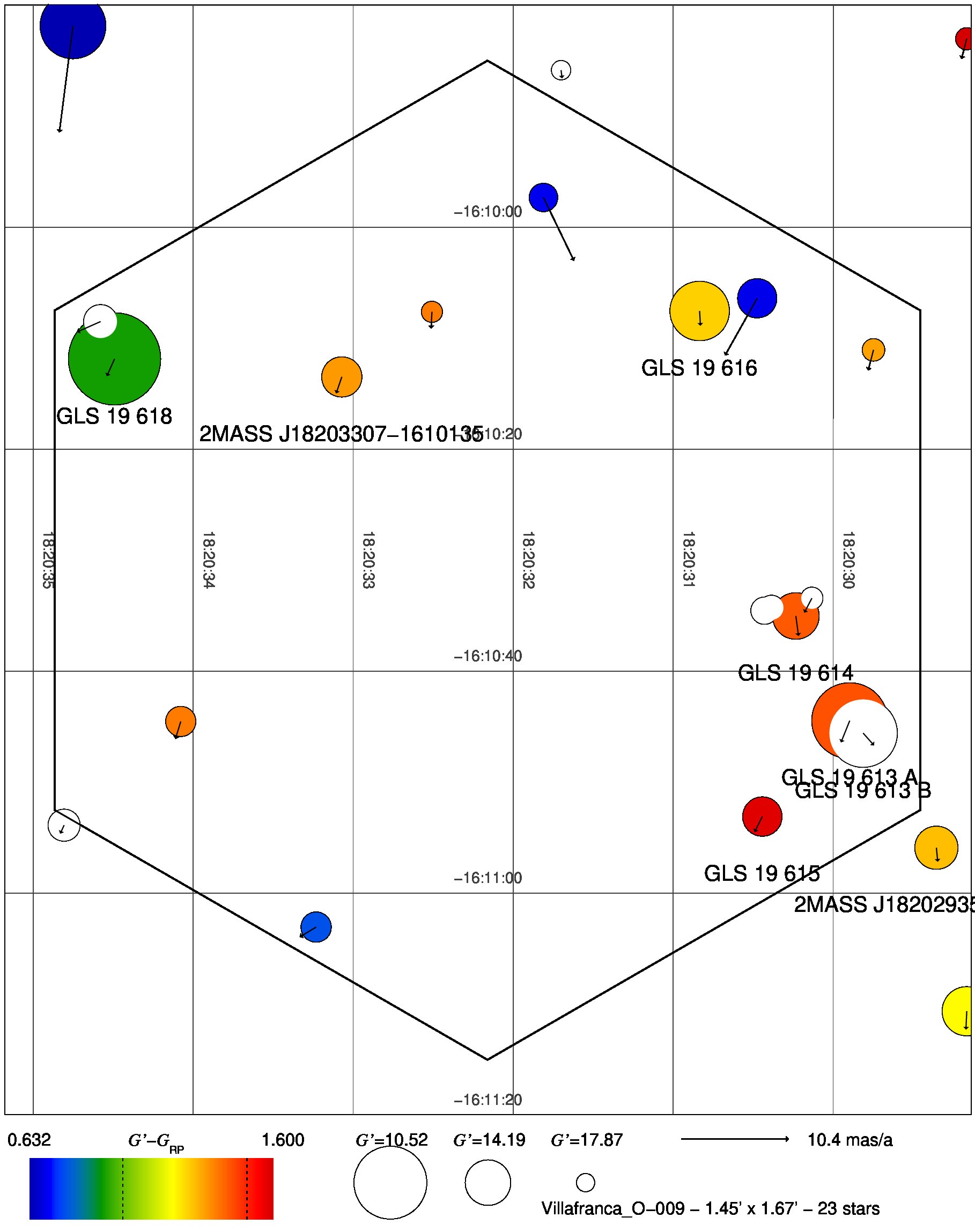}}\,%
                  \includegraphics*[width=0.380\linewidth]{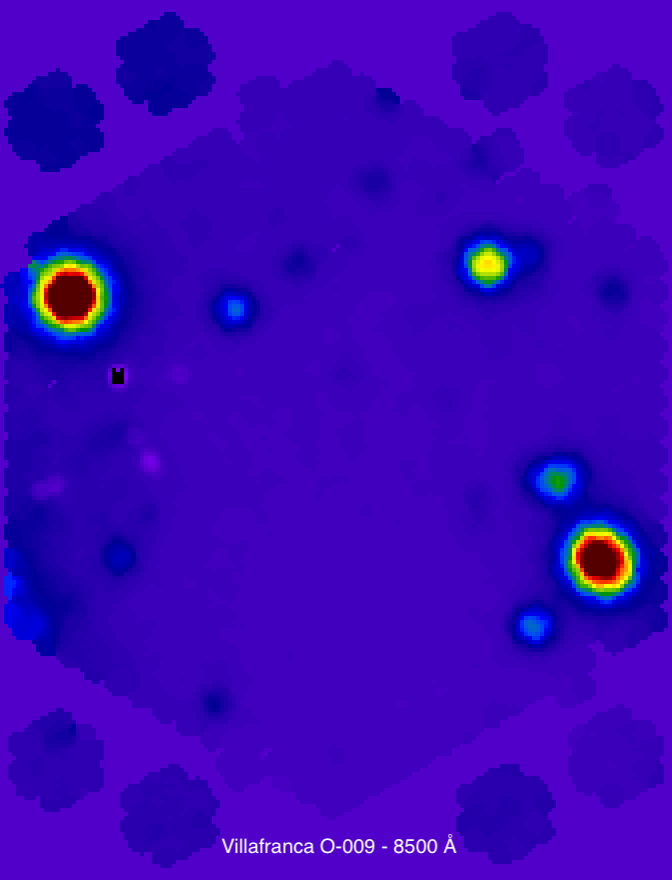}\,%
                  \includegraphics*[width=0.380\linewidth]{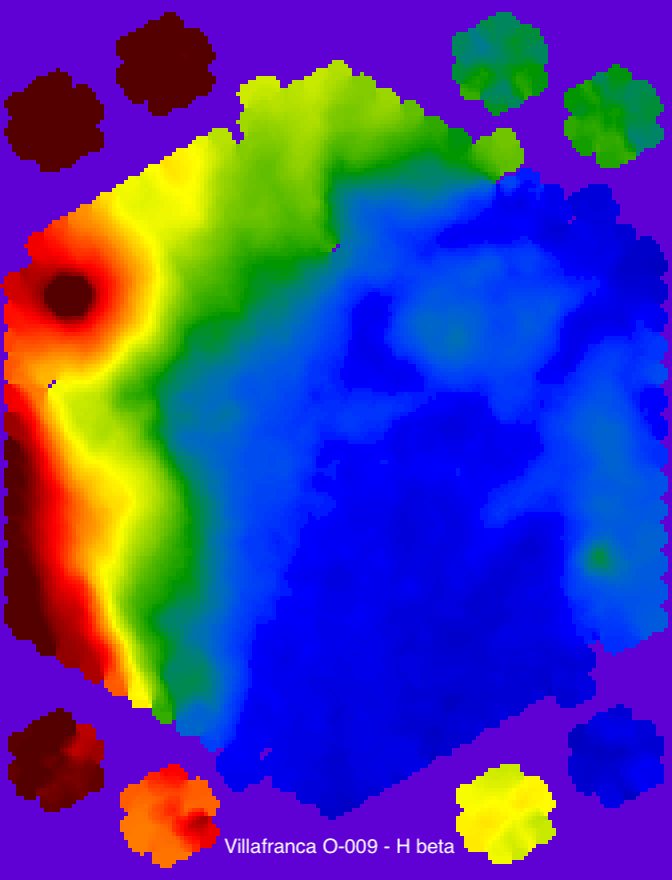}\,%
             }
 \caption{Sample fields. The left panels are \textit{Gaia} charts and the other two are cube slices at different wavelengths.}
 \label{fig2}   
\end{figure}
\end{landscape}

\addtocounter{figure}{-1}

\begin{landscape}
\begin{figure}
 \centerline{
  \hspace{-3.2cm}
  \raisebox{-8mm}{\includegraphics*[width=0.410\linewidth]{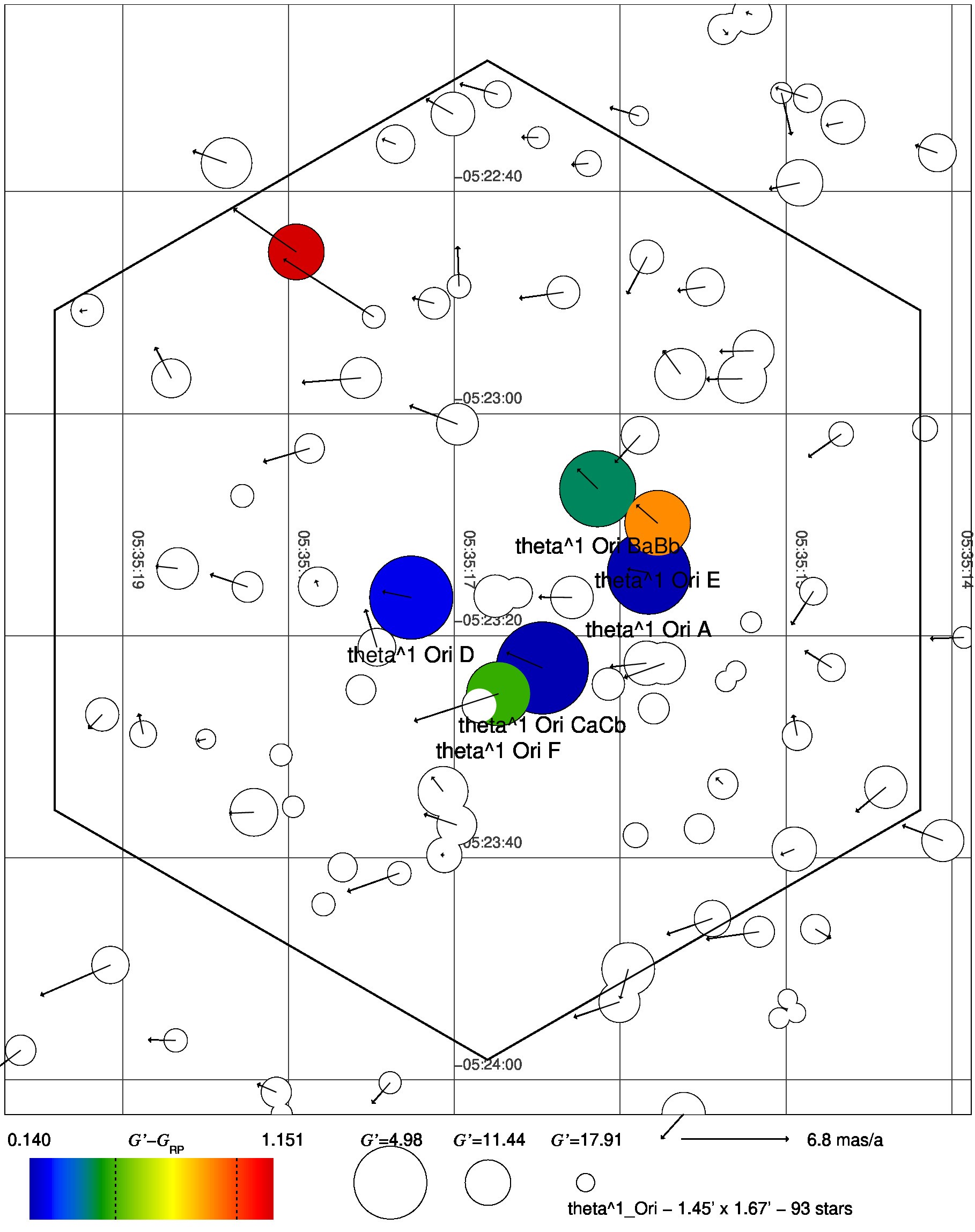}}\,%
                  \includegraphics*[width=0.380\linewidth]{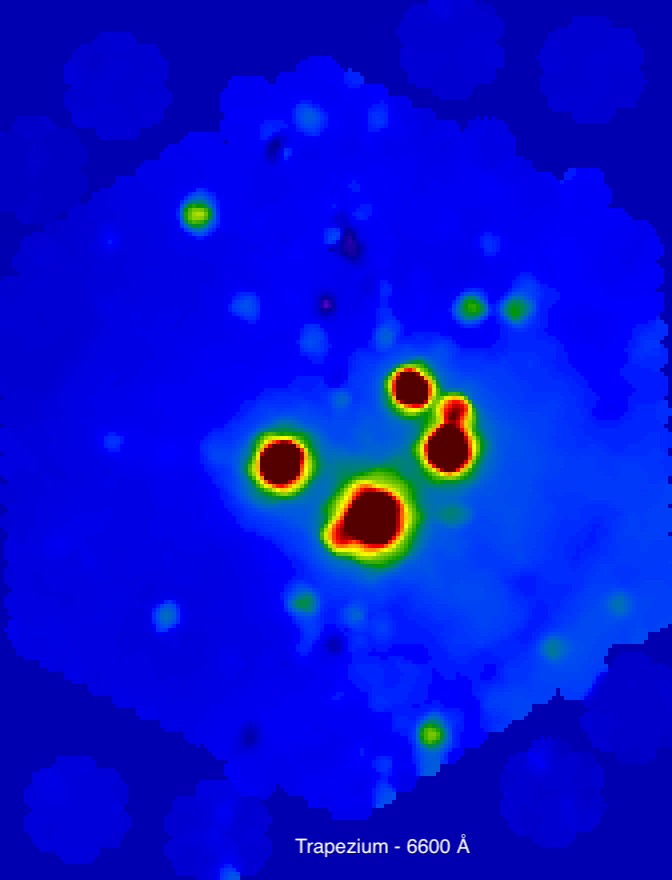}\,%
                  \includegraphics*[width=0.380\linewidth]{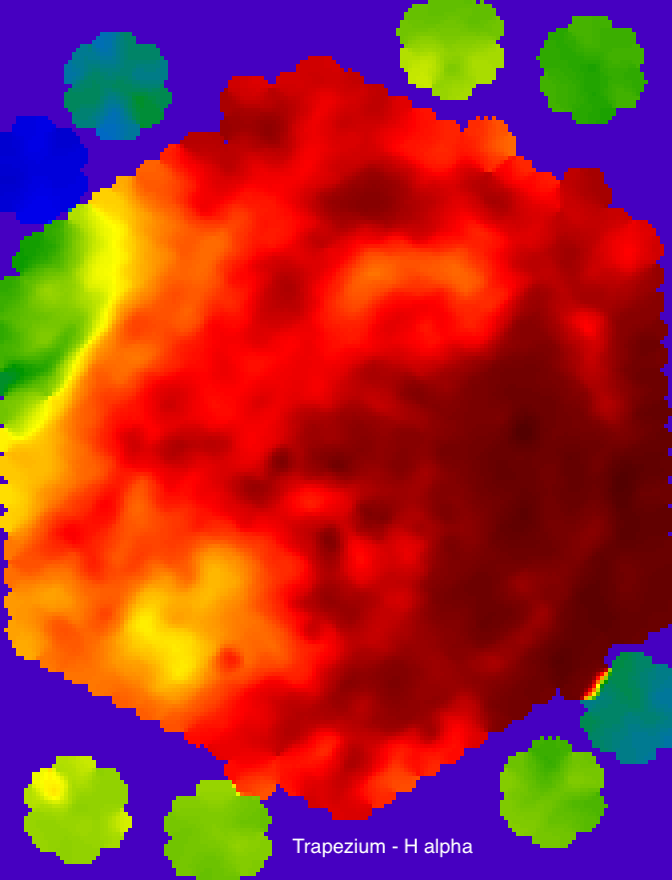}\,%
             }
 \caption{(Continued).}
\end{figure}
\end{landscape}

\addtocounter{figure}{-1}

\begin{landscape}
\begin{figure}
 \centerline{
  \hspace{-3.2cm}
  \raisebox{-8mm}{\includegraphics*[width=0.410\linewidth]{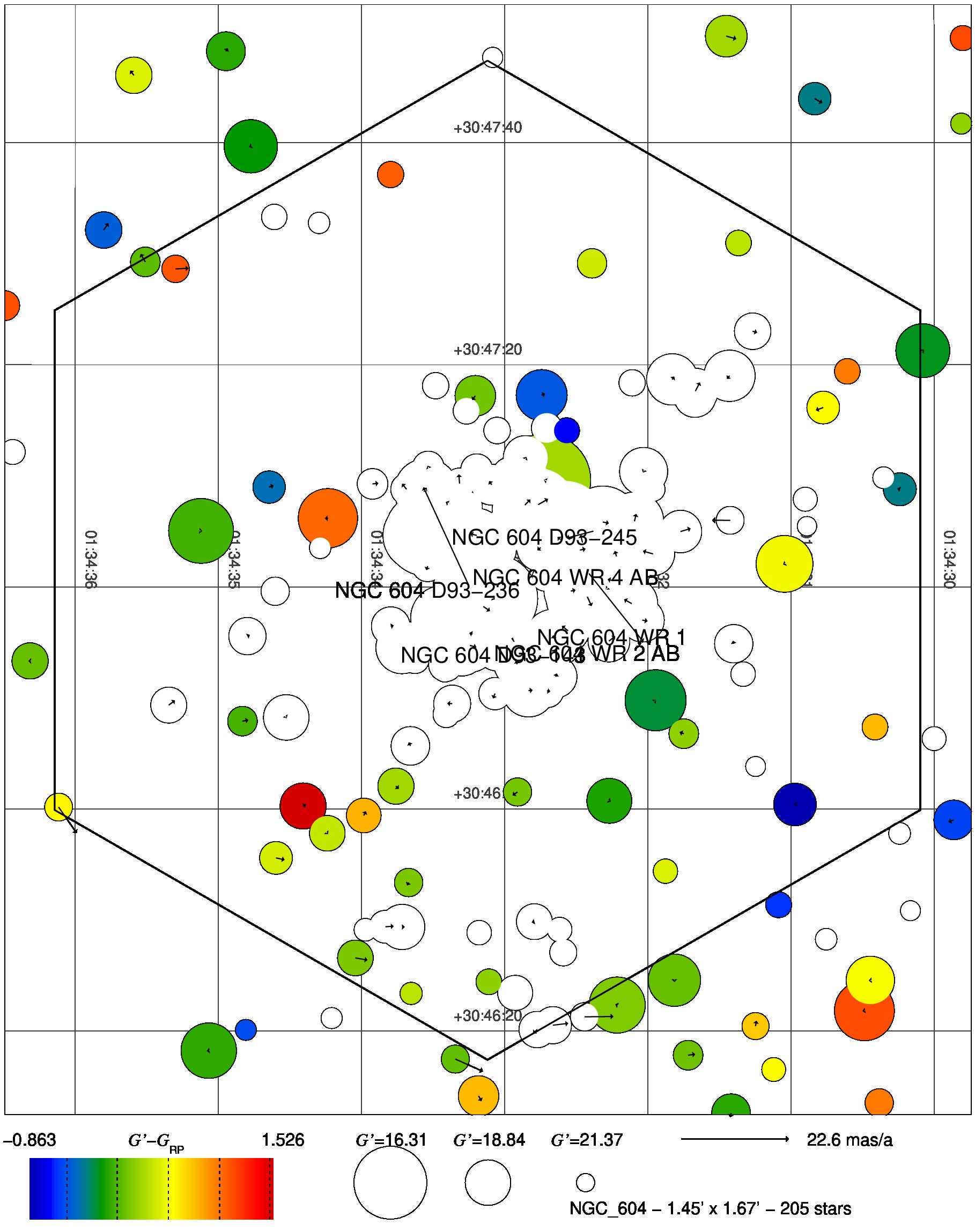}}\,%
                  \includegraphics*[width=0.380\linewidth]{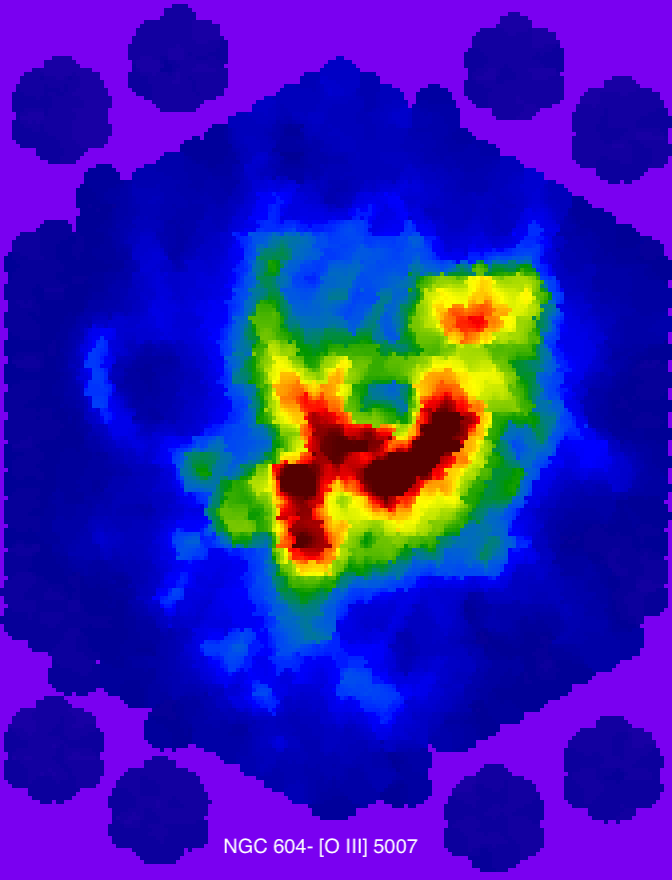}\,%
                  \includegraphics*[width=0.380\linewidth]{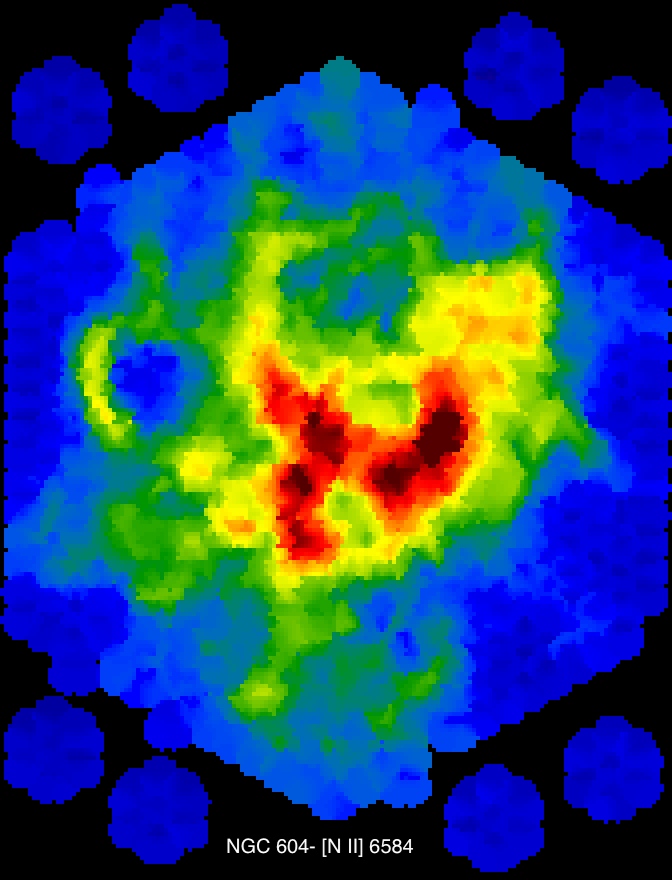}\,%
             }
 \caption{(Continued).}
\end{figure}
\end{landscape}

\addtocounter{figure}{-1}

\begin{landscape}
\begin{figure}
 \centerline{
  \hspace{-3.2cm}
  \raisebox{-8mm}{\includegraphics*[width=0.410\linewidth]{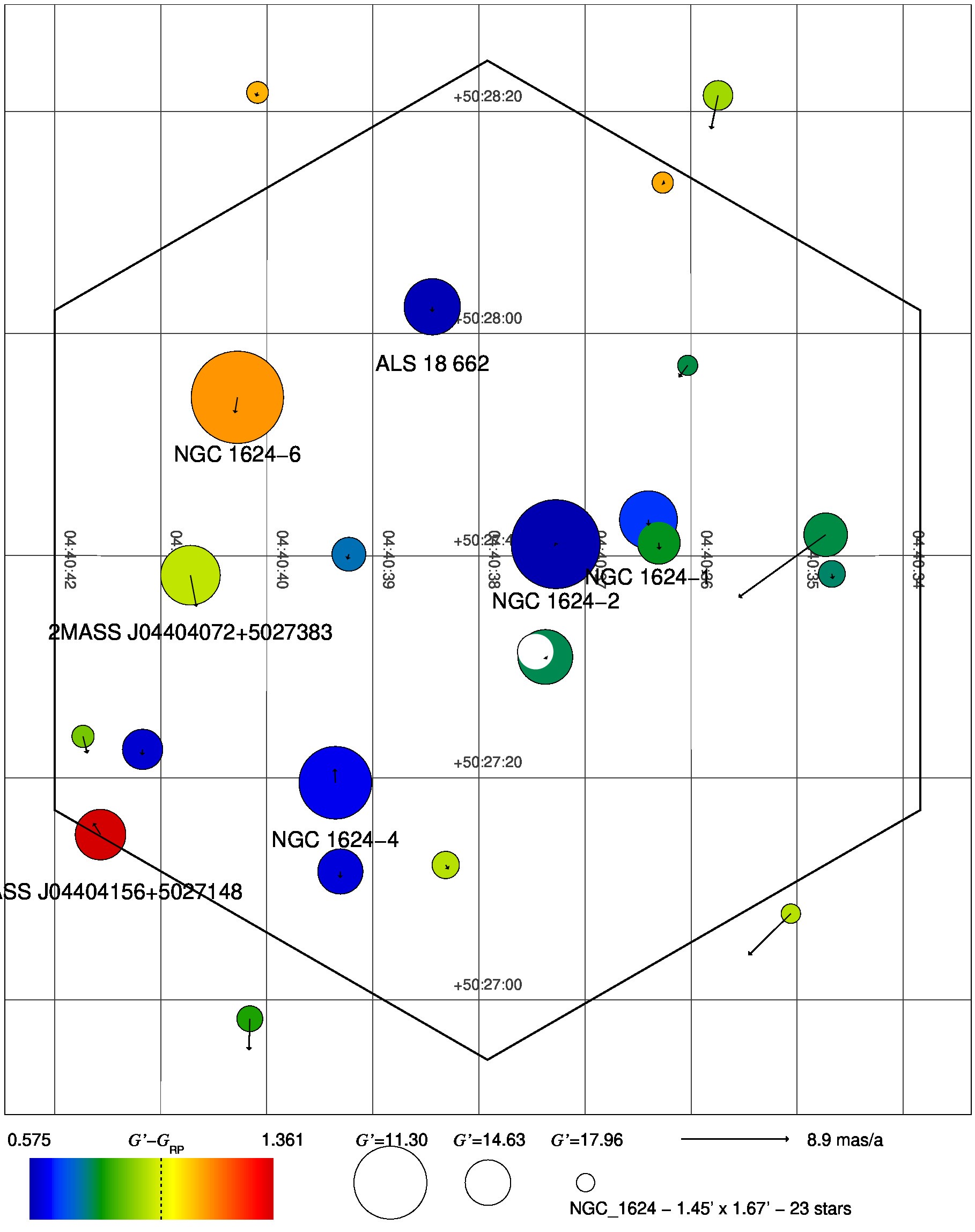}}\,%
                  \includegraphics*[width=0.380\linewidth]{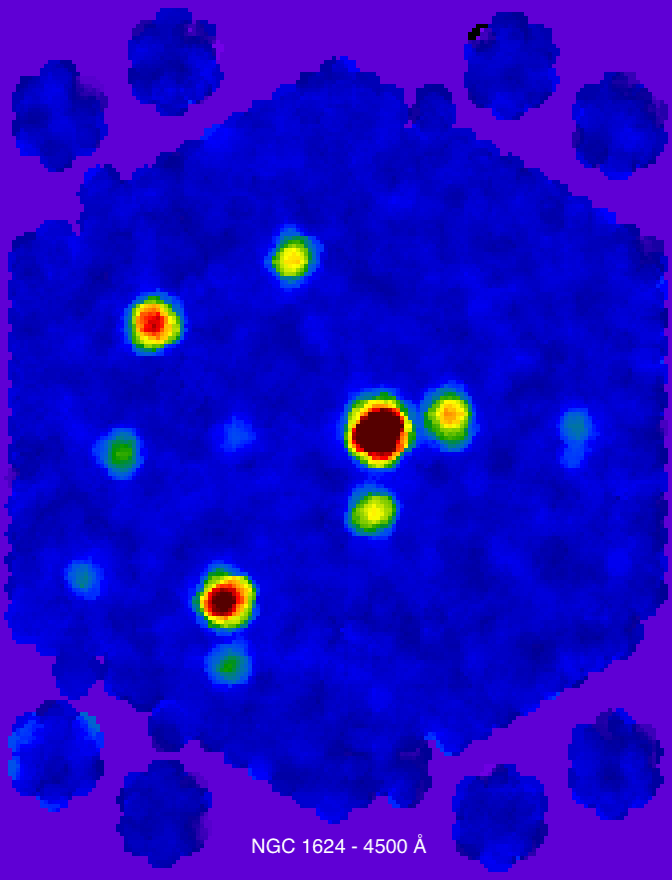}\,%
                  \includegraphics*[width=0.380\linewidth]{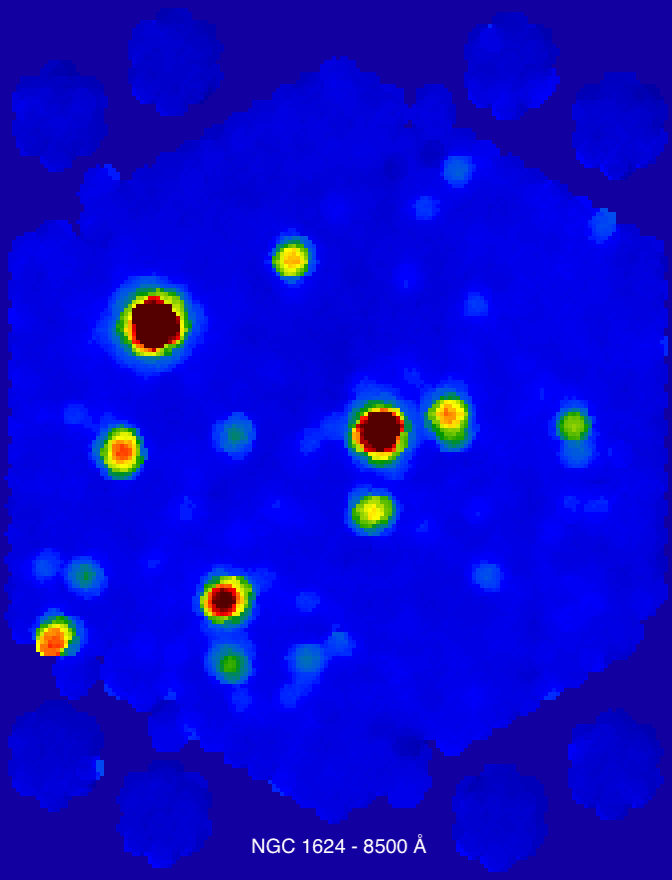}\,%
             }
 \caption{(Continued).}
\end{figure}
\end{landscape}

\addtocounter{figure}{-1}

\begin{landscape}
\begin{figure}
 \centerline{
  \hspace{-3.2cm}
  \raisebox{-8mm}{\includegraphics*[width=0.410\linewidth]{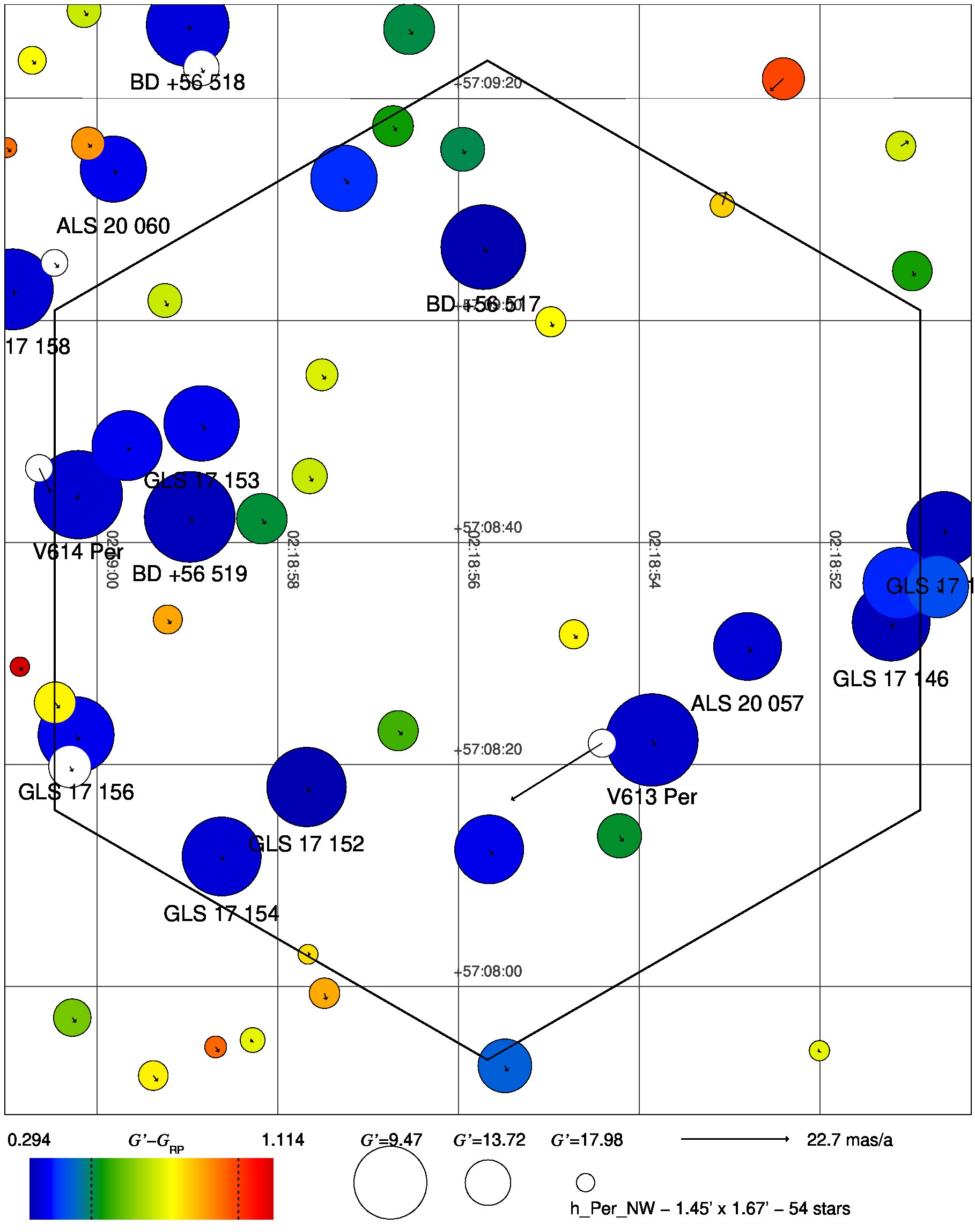}}\,%
                  \includegraphics*[width=0.380\linewidth]{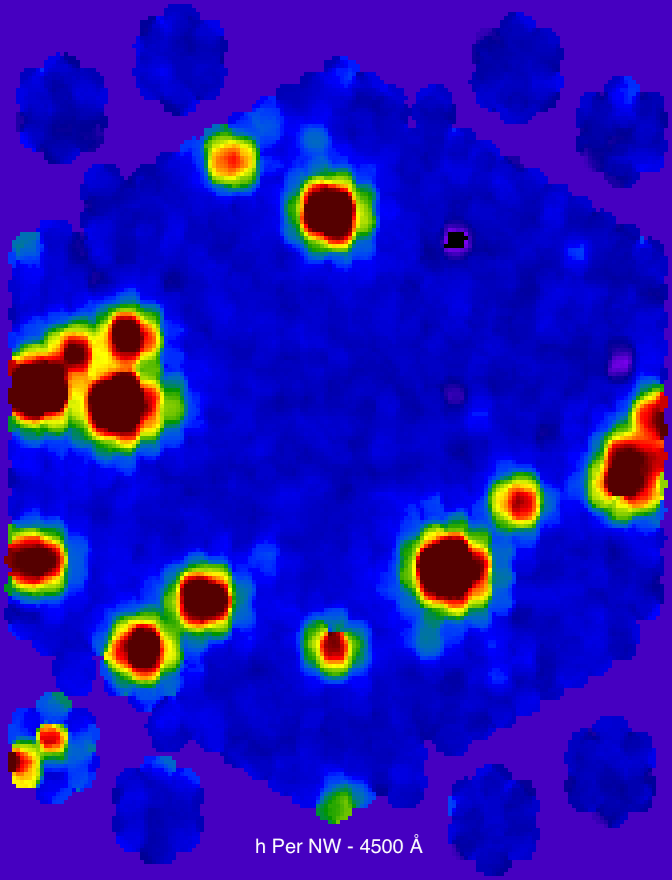}\,%
                  \includegraphics*[width=0.380\linewidth]{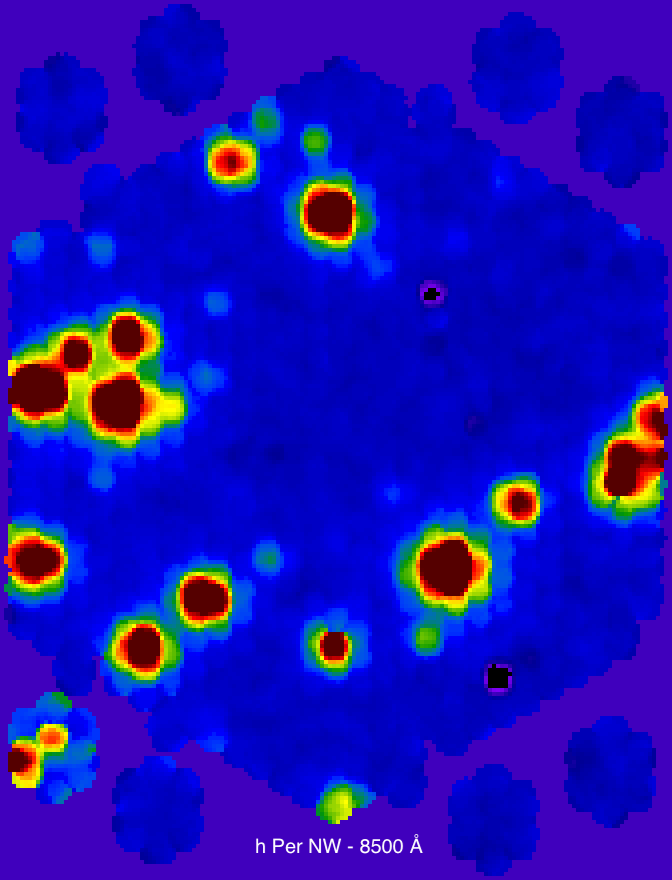}\,%
             }
 \caption{(Continued).}
\end{figure}
\end{landscape}

\addtocounter{figure}{-1}

\begin{landscape}
\begin{figure}
 \centerline{
  \hspace{-3.2cm}
  \raisebox{-8mm}{\includegraphics*[width=0.410\linewidth]{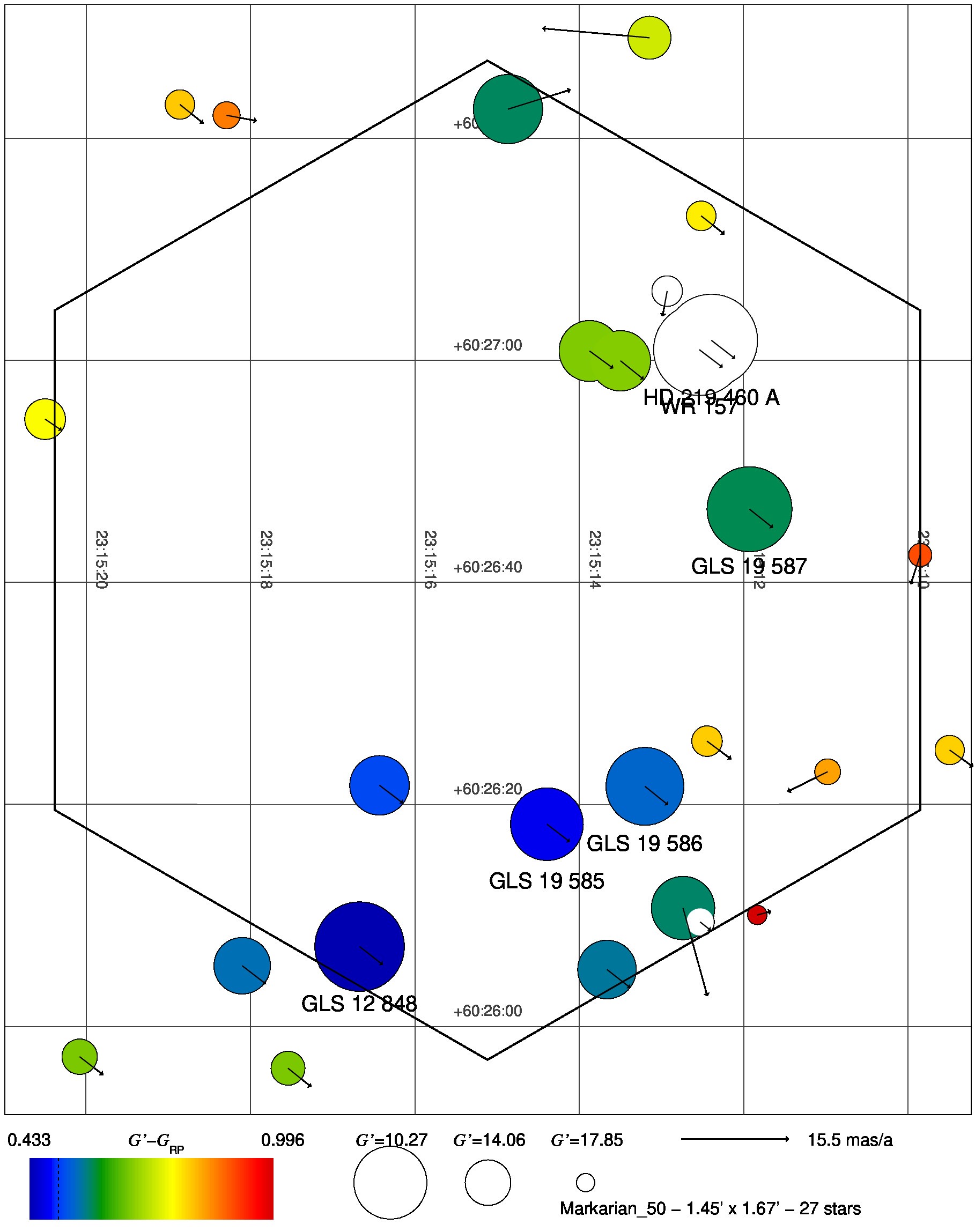}}\,%
                  \includegraphics*[width=0.380\linewidth]{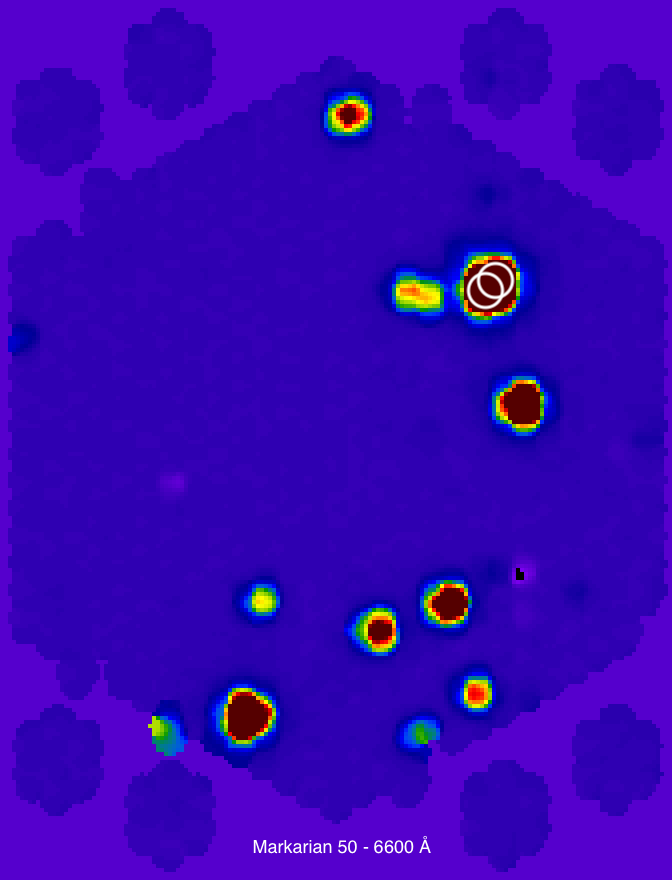}\,%
                  \includegraphics*[width=0.380\linewidth]{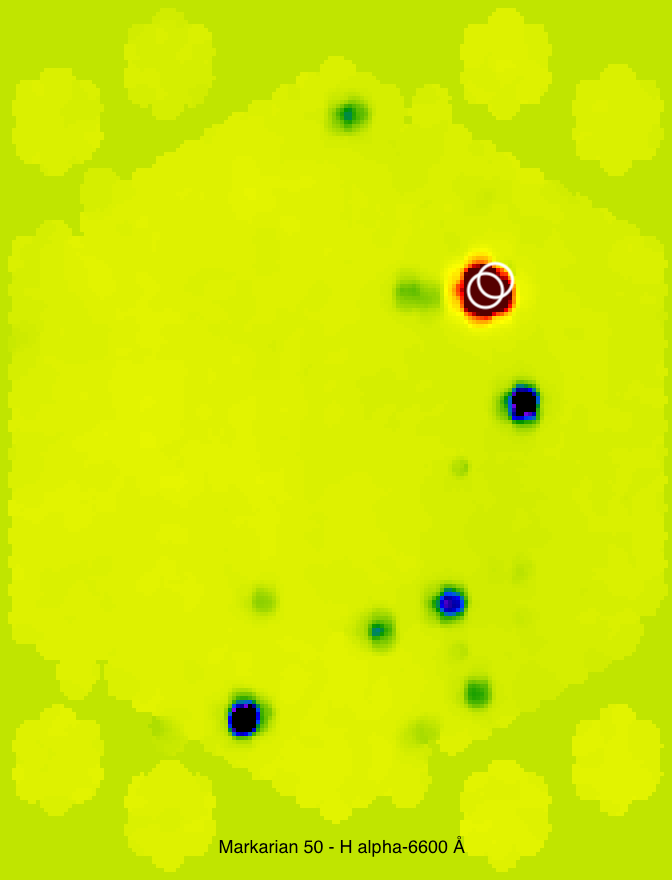}\,%
             }
 \caption{(Continued).}
\end{figure}
\end{landscape}

\addtocounter{figure}{-1}

\begin{landscape}
\begin{figure}
 \centerline{
  \hspace{-3.2cm}
  \raisebox{-8mm}{\includegraphics*[width=0.410\linewidth]{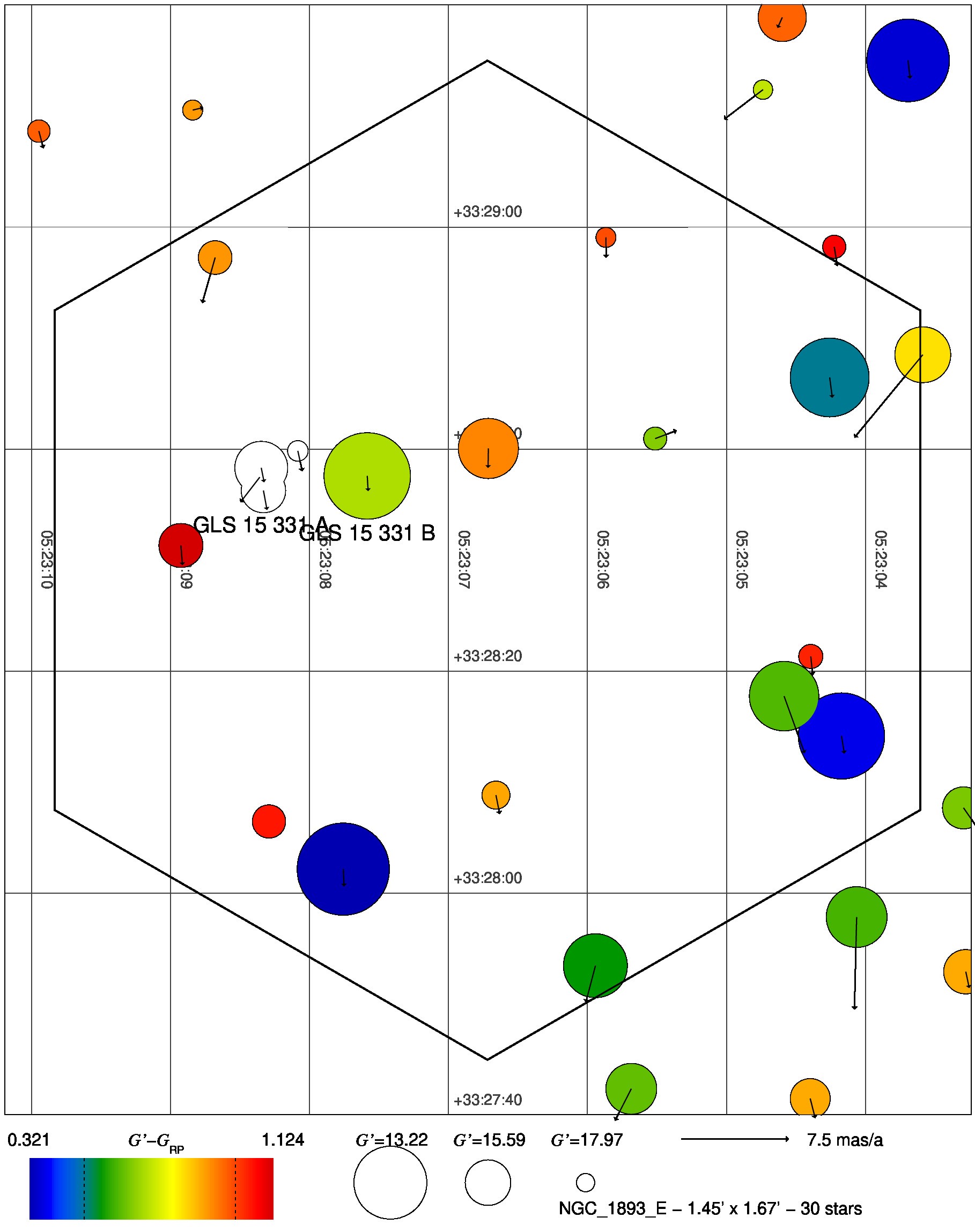}}\,%
                  \includegraphics*[width=0.380\linewidth]{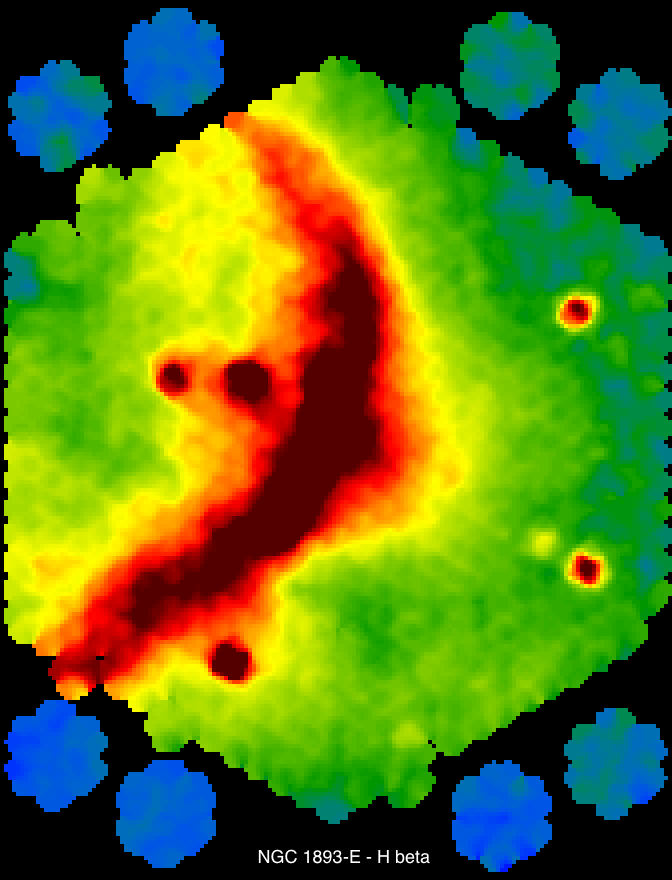}\,%
                  \includegraphics*[width=0.380\linewidth]{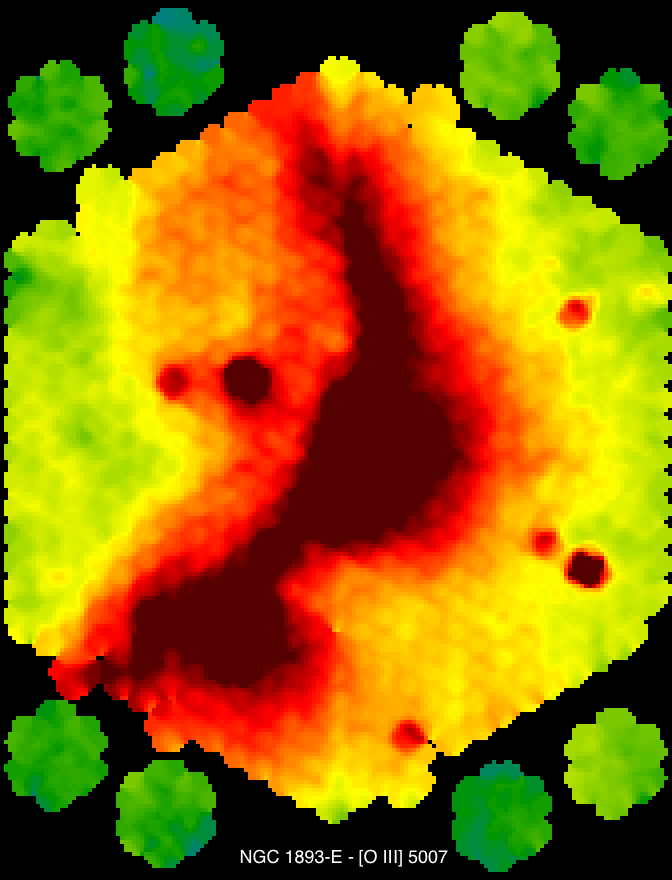}\,%
             }
 \caption{(Continued).}
\end{figure}
\end{landscape}

\addtocounter{figure}{-1}

\begin{landscape}
\begin{figure}
 \centerline{
  \hspace{-3.2cm}
  \raisebox{-8mm}{\includegraphics*[width=0.410\linewidth]{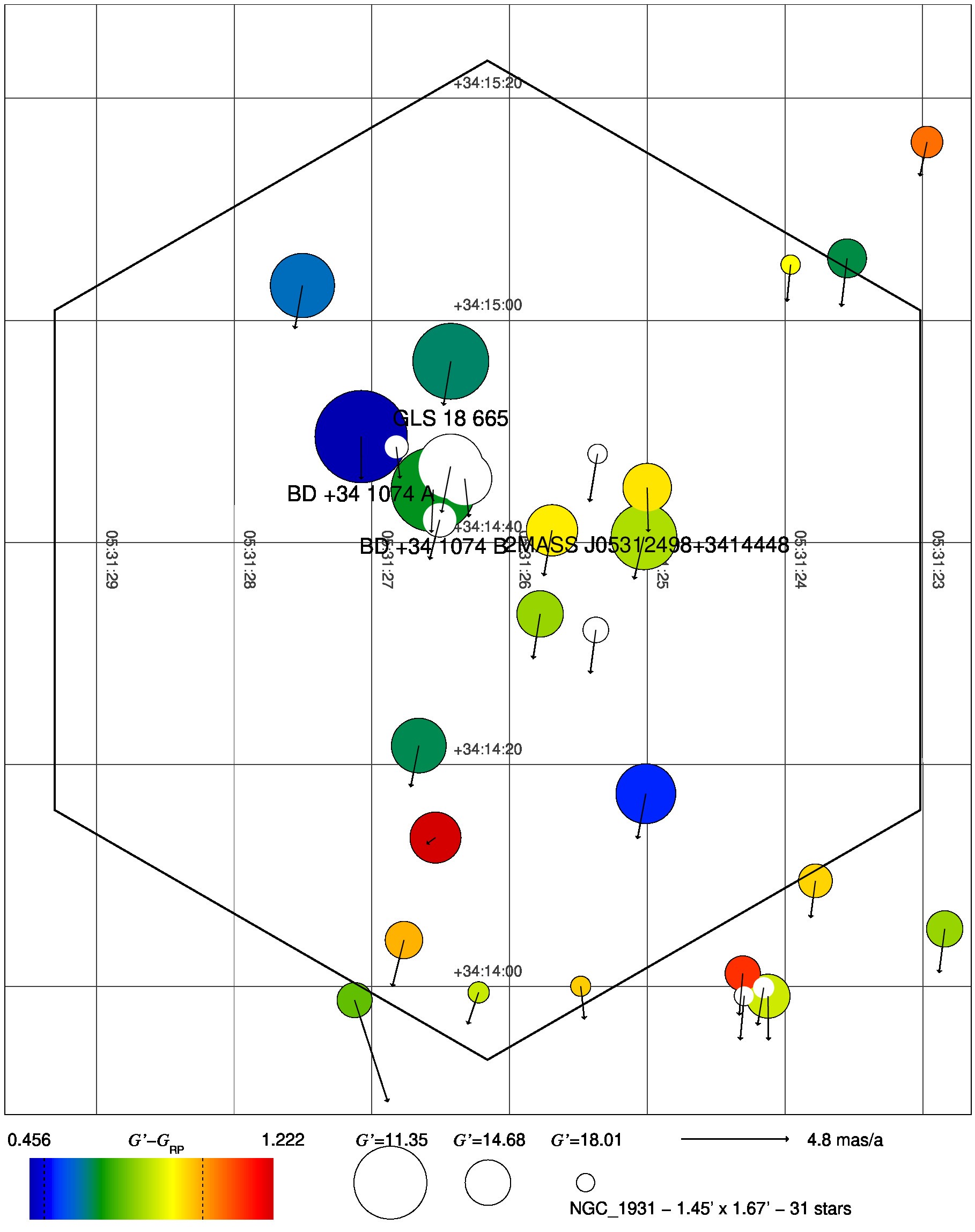}}\,%
                  \includegraphics*[width=0.380\linewidth]{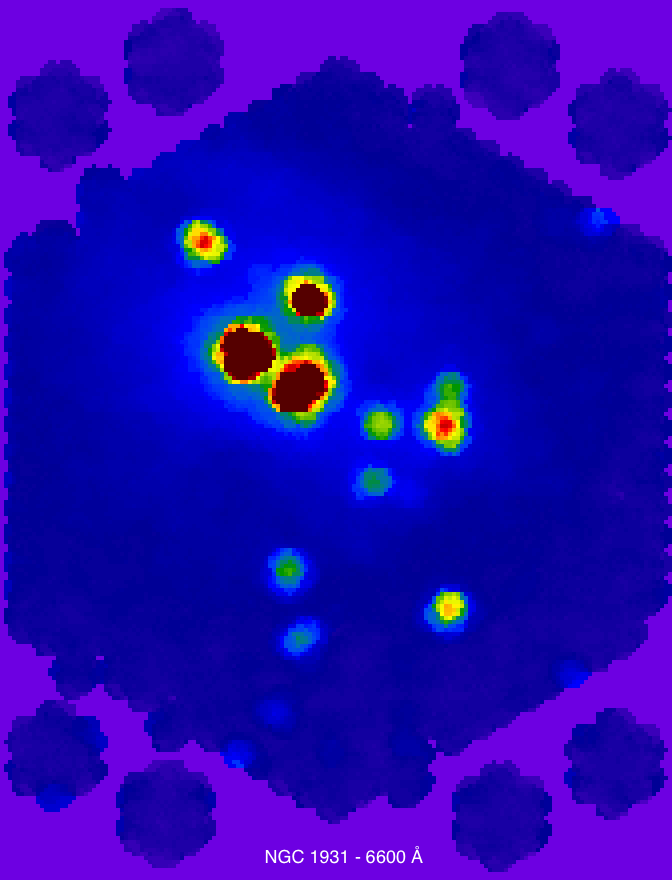}\,%
                  \includegraphics*[width=0.380\linewidth]{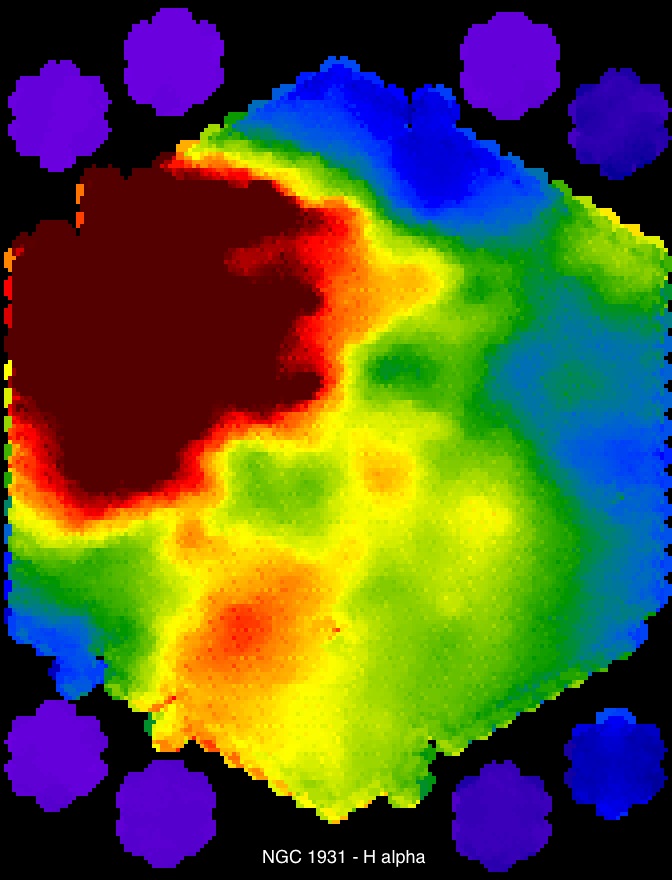}\,%
             }
 \caption{(Continued).}
\end{figure}
\end{landscape}

\addtocounter{figure}{-1}

\begin{landscape}
\begin{figure}
 \centerline{
  \hspace{-3.2cm}
  \raisebox{-8mm}{\includegraphics*[width=0.410\linewidth]{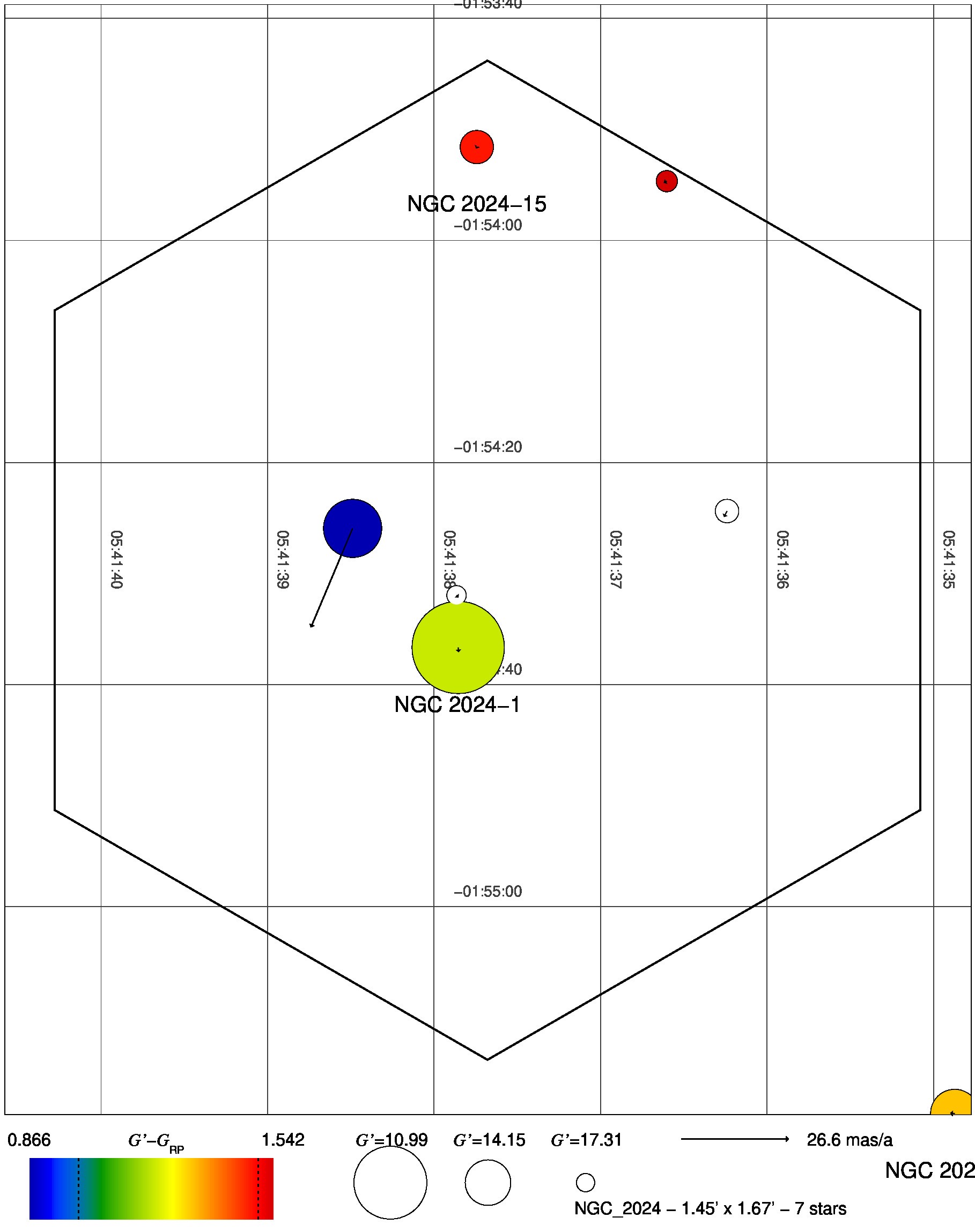}}\,%
                  \includegraphics*[width=0.380\linewidth]{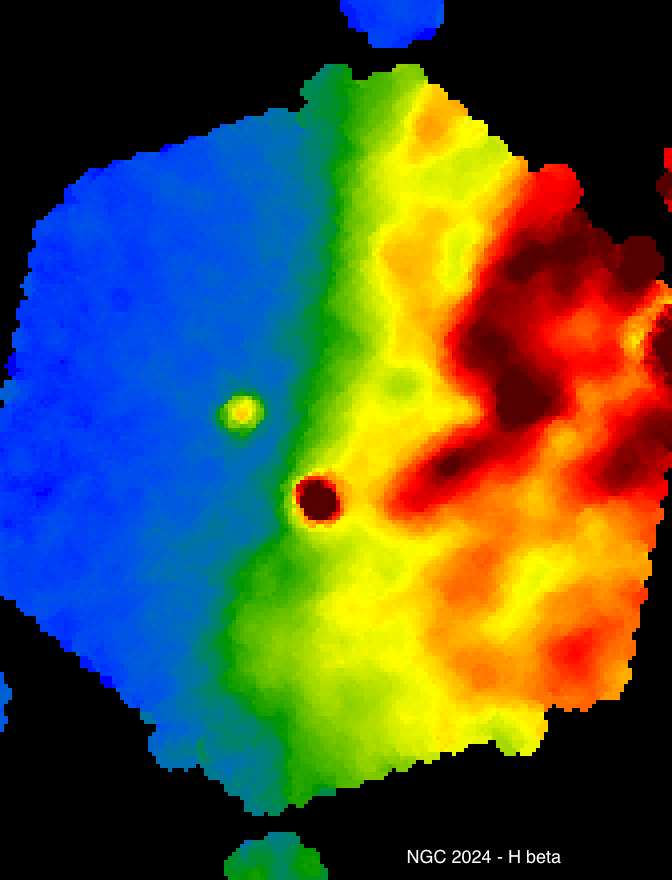}\,%
                  \includegraphics*[width=0.380\linewidth]{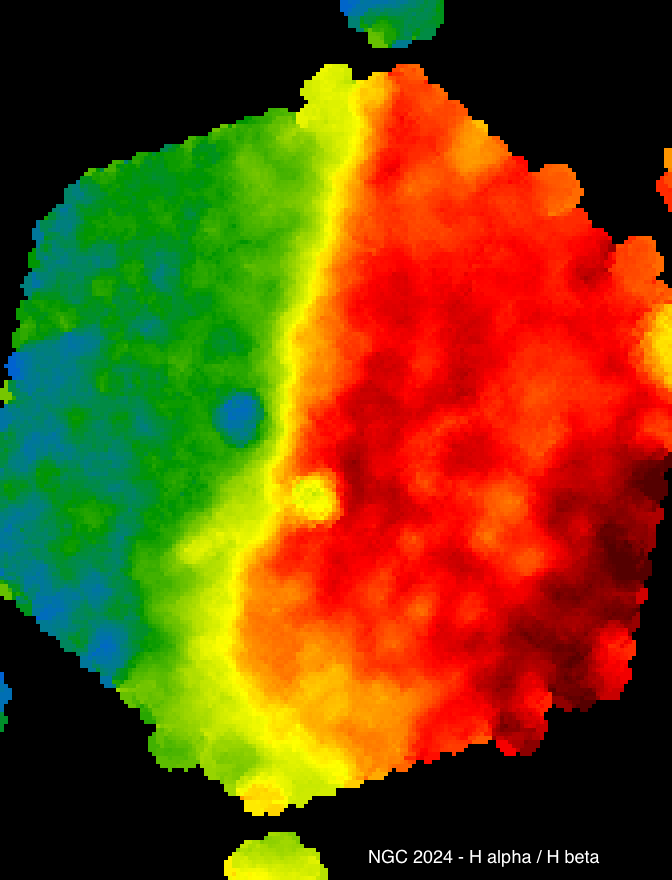}\,%
             }
 \caption{(Continued).}
\end{figure}
\end{landscape}

\addtocounter{figure}{-1}

\begin{landscape}
\begin{figure}
 \centerline{
  \hspace{-3.2cm}
  \raisebox{-8mm}{\includegraphics*[width=0.410\linewidth]{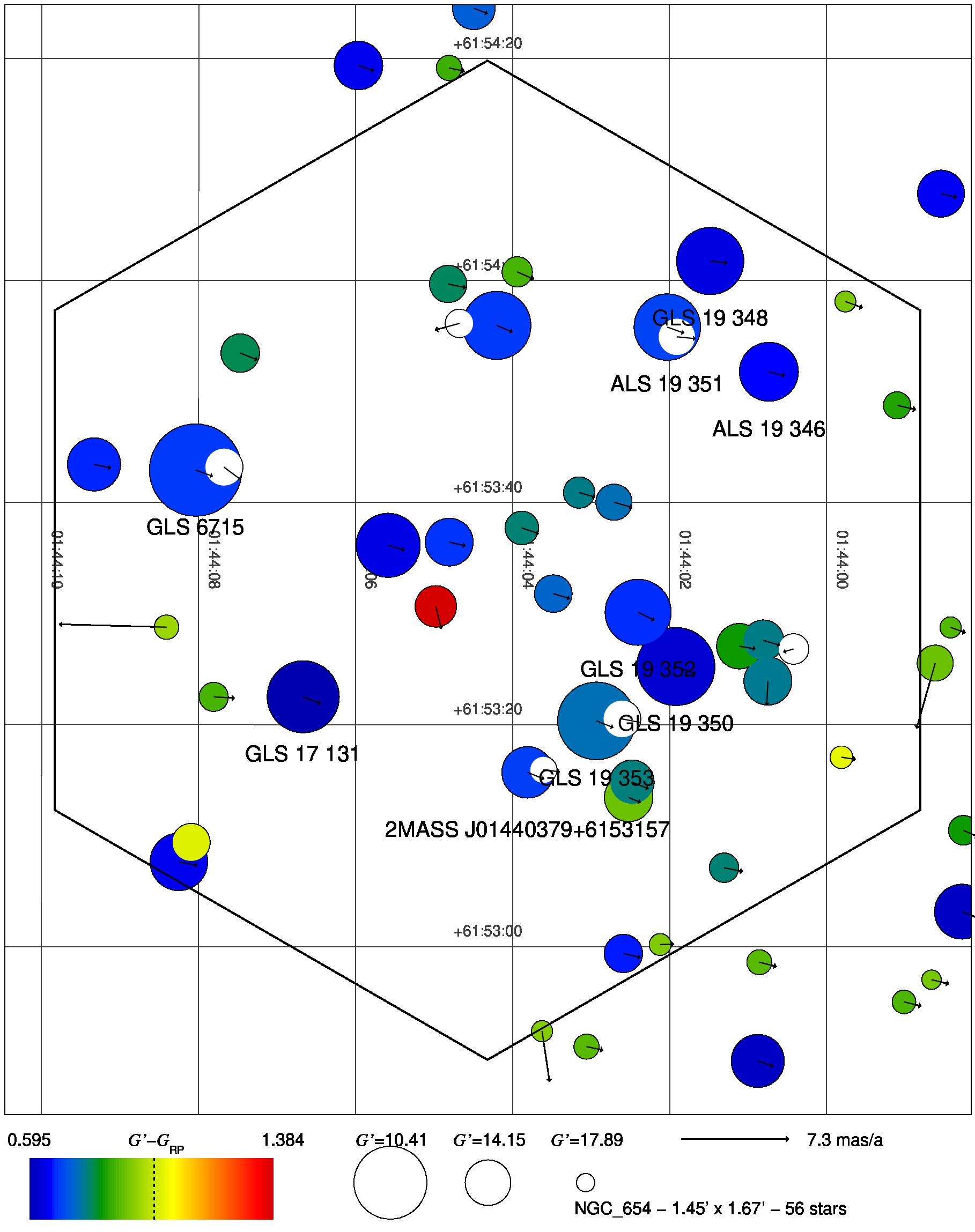}}\,%
                  \includegraphics*[width=0.380\linewidth]{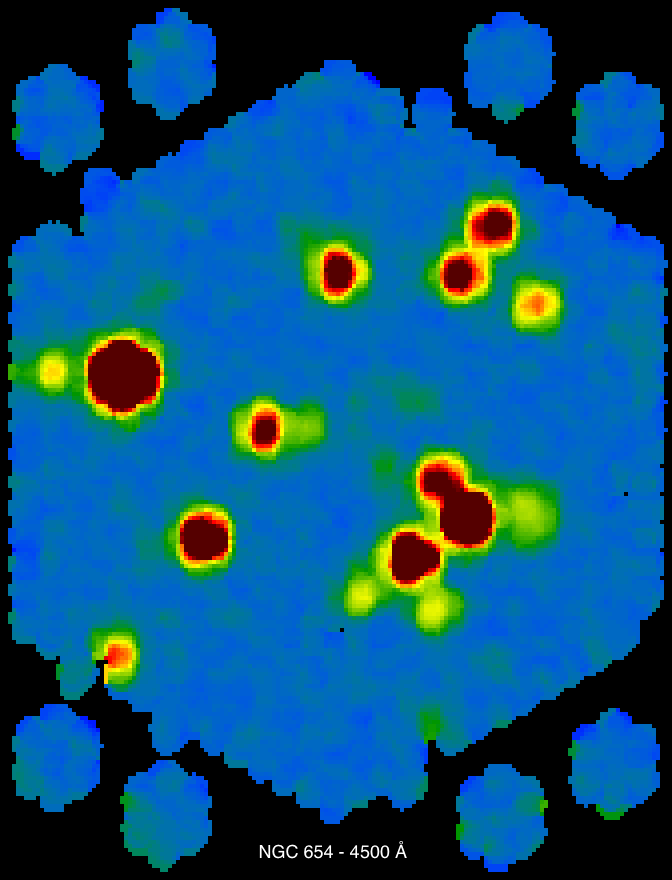}\,%
                  \includegraphics*[width=0.380\linewidth]{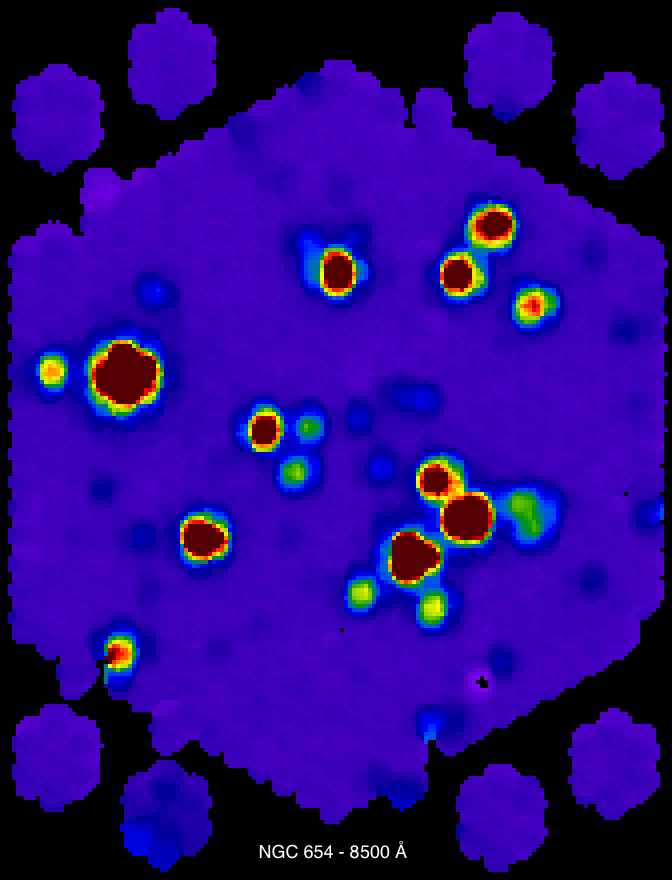}\,%
             }
 \caption{(Continued).}
\end{figure}
\end{landscape}

\end{document}

%% file: mlaeff.tex
\def\hoy{\number\day \space de \space\ifcase\month\or
 Enero\or Febrero\or Marzo\or Abril\or Mayo\or Junio\or
 Julio\or Agosto\or Septiembre\or Octubre\or Noviembre\or Diciembre\fi
 \space de \number\year}
\def\ii/{\'{\i}}
\def\cion/{ci\'on}
\def\cao/{\c c\~ao}
\def\arcsec{\hbox{$^{\prime\prime}$}}
\def\utw{\smash{\rlap{\lower5pt\hbox{$\sim$}}}}
\def\udtw{\smash{\rlap{\lower6pt\hbox{$\approx$}}}}

\def\farcs{\hbox{$.\!\!^{\prime\prime}$}}

\def\tens#1{\ifmmode\mathchoice{\mbox{$\sf\displaystyle#1$}}
{\mbox{$\sf\textstyle#1$}}
{\mbox{$\sf\scriptstyle#1$}}
{\mbox{$\sf\scriptscriptstyle#1$}}\else
\hbox{$\sf\textstyle#1$}\fi}
\def\vec#1{\ifmmode\mathchoice{\mbox{\boldmath$\displaystyle#1$}}
{\mbox{\boldmath$\textstyle#1$}}
{\mbox{\boldmath$\scriptstyle#1$}}
{\mbox{\boldmath$\scriptscriptstyle#1$}}\else
\hbox{\boldmath$\textstyle#1$}\fi}
\def\bbbc{{\mathchoice {\setbox0=\hbox{$\displaystyle\rm C$}\hbox{\hbox
to0pt{\kern0.4\wd0\vrule height0.9\ht0\hss}\box0}}
{\setbox0=\hbox{$\textstyle\rm C$}\hbox{\hbox
to0pt{\kern0.4\wd0\vrule height0.9\ht0\hss}\box0}}
{\setbox0=\hbox{$\scriptstyle\rm C$}\hbox{\hbox
to0pt{\kern0.4\wd0\vrule height0.9\ht0\hss}\box0}}
{\setbox0=\hbox{$\scriptscriptstyle\rm C$}\hbox{\hbox
to0pt{\kern0.4\wd0\vrule height0.9\ht0\hss}\box0}}}}
\def\bbbq{{\mathchoice {\setbox0=\hbox{$\displaystyle\rm
Q$}\hbox{\raise
0.15\ht0\hbox to0pt{\kern0.4\wd0\vrule height0.8\ht0\hss}\box0}}
{\setbox0=\hbox{$\textstyle\rm Q$}\hbox{\raise
0.15\ht0\hbox to0pt{\kern0.4\wd0\vrule height0.8\ht0\hss}\box0}}
{\setbox0=\hbox{$\scriptstyle\rm Q$}\hbox{\raise
0.15\ht0\hbox to0pt{\kern0.4\wd0\vrule height0.7\ht0\hss}\box0}}
{\setbox0=\hbox{$\scriptscriptstyle\rm Q$}\hbox{\raise
0.15\ht0\hbox to0pt{\kern0.4\wd0\vrule height0.7\ht0\hss}\box0}}}}
\def\bbbt{{\mathchoice {\setbox0=\hbox{$\displaystyle\rm
T$}\hbox{\hbox to0pt{\kern0.3\wd0\vrule height0.9\ht0\hss}\box0}}
{\setbox0=\hbox{$\textstyle\rm T$}\hbox{\hbox
to0pt{\kern0.3\wd0\vrule height0.9\ht0\hss}\box0}}
{\setbox0=\hbox{$\scriptstyle\rm T$}\hbox{\hbox
to0pt{\kern0.3\wd0\vrule height0.9\ht0\hss}\box0}}
{\setbox0=\hbox{$\scriptscriptstyle\rm T$}\hbox{\hbox
to0pt{\kern0.3\wd0\vrule height0.9\ht0\hss}\box0}}}}
\def\bbbs{{\mathchoice
{\setbox0=\hbox{$\displaystyle     \rm S$}\hbox{\raise0.5\ht0\hbox
to0pt{\kern0.35\wd0\vrule height0.45\ht0\hss}\hbox
to0pt{\kern0.55\wd0\vrule height0.5\ht0\hss}\box0}}
{\setbox0=\hbox{$\textstyle        \rm S$}\hbox{\raise0.5\ht0\hbox
to0pt{\kern0.35\wd0\vrule height0.45\ht0\hss}\hbox
to0pt{\kern0.55\wd0\vrule height0.5\ht0\hss}\box0}}
{\setbox0=\hbox{$\scriptstyle      \rm S$}\hbox{\raise0.5\ht0\hbox
to0pt{\kern0.35\wd0\vrule height0.45\ht0\hss}\raise0.05\ht0\hbox
to0pt{\kern0.5\wd0\vrule height0.45\ht0\hss}\box0}}
{\setbox0=\hbox{$\scriptscriptstyle\rm S$}\hbox{\raise0.5\ht0\hbox
to0pt{\kern0.4\wd0\vrule height0.45\ht0\hss}\raise0.05\ht0\hbox
to0pt{\kern0.55\wd0\vrule height0.45\ht0\hss}\box0}}}}
\def\bbbz{{\mathchoice {\hbox{$\sf\textstyle Z\kern-0.4em Z$}}
{\hbox{$\sf\textstyle Z\kern-0.4em Z$}}
{\hbox{$\sf\scriptstyle Z\kern-0.3em Z$}}
{\hbox{$\sf\scriptscriptstyle Z\kern-0.2em Z$}}}}
\def\diameter{{\ifmmode\mathchoice
{\ooalign{\hfil\hbox{$\displaystyle/$}\hfil\crcr
{\hbox{$\displaystyle\mathchar"20D$}}}}
{\ooalign{\hfil\hbox{$\textstyle/$}\hfil\crcr
{\hbox{$\textstyle\mathchar"20D$}}}}
{\ooalign{\hfil\hbox{$\scriptstyle/$}\hfil\crcr
{\hbox{$\scriptstyle\mathchar"20D$}}}}
{\ooalign{\hfil\hbox{$\scriptscriptstyle/$}\hfil\crcr
{\hbox{$\scriptscriptstyle\mathchar"20D$}}}}
\else{\ooalign{\hfil/\hfil\crcr\mathhexbox20D}}%
\fi}}
\def\sq{\ifmmode\squareforqed\else{\unskip\nobreak\hfil
\penalty50\hskip1em\null\nobreak\hfil\squareforqed
\parfillskip=0pt\finalhyphendemerits=0\endgraf}\fi}
\def\squareforqed{\hbox{\rlap{$\sqcap$}$\sqcup$}}